\documentclass[review,amsmath,amssymb,nofootinbib,superscriptaddress,a4paper]{revtex4}
\usepackage{array,mathtools,amssymb,booktabs}
\usepackage{hyperref}
\usepackage[english]{layout}
\usepackage[english]{babel}
\usepackage{bm}
\usepackage{graphicx}
\usepackage[left=3cm, right=3cm, top=3cm, bottom=2cm]{geometry}
\usepackage{hhline}

\begin{document}

\title{Light Meson Masses using AdS/QCD modified Soft Wall Model}
\author{Santiago Cort\'es}
\email{js.cortes125@uniandes.edu.co}
\affiliation{Departamento de F\'{\i}sica, Univ. de Los  Andes, 111711 Bogot\'a, Colombia.}
\author{Miguel \'Angel Mart\'{\i}n Contreras}
\email{ma.martin41@uniandes.edu.co}
\affiliation{Departamento de F\'{\i}sica, Univ. de Los  Andes, 111711 Bogot\'a, Colombia.}
\author{Jos\'e Rolando Rold\'an}
\email{jroldan@uniandes.edu.co}
\affiliation{Departamento de F\'{\i}sica, Univ. de Los  Andes, 111711 Bogot\'a, Colombia.}
\date{\today}

\begin{abstract}
We analyze light vector and scalar meson mass spectra using a novel approach where a modified soft wall model with a UV-cutoff is considered. Including this cutoff introduces an extra energy scale. For this model, we found that the masses for the scalar and vector spectra are well fitted within a very small RMS error for 14 of these states, with non-linear trajectories given by two common parameters, the UV locus $z_0$ and the quadratic dilaton profile slope $\kappa$.  We concluded that in this model the $f_{0}(500)$ scalar resonance cannot be fitted holographycally as a $q\overline{q}$ state since we could not find a trajectory that included this pole. This result is in agreement with the most recent phenomenological and theoretical methods.

%\vspace*{0.5cm}
%\pacs{ 11.10.Wx, % Finite temperature field theory
%12.39.Fe, % Chiral lagrangians
%11.15.Pg,  % Expansions for large numbers of components (e.g., 1/Nc expansions)
%11.30.Rd % Chiral symmetries
%}
\end{abstract}

\maketitle



%Gauge/string duality, 11.25.Tq
%properties of baryons, 14.20.-c mesons, 14.40.-n
%Light mesons, 14.40.Be
%Quantum chromodynamics, 12.38.-t

\section{Introduction} 

The idea of using the AdS/CFT correspondence \cite{Maldacena:1997re,Ramallo:2013bua} to describe nonperturbative QCD-like phenomena has given insights to explore the strong interactions at strong coupling, unreachable by regular QFT methods. One possibility is considering gravity models in a given space that holographically generate low-energy QCD theories living on the conformal flat boundary. This proposal is called top-down. The second scenario considers the opposite: starting from well known properties derived from a 4-dimensional QCD, one tries to look out for a gravity theory which is its holographic dual; this is the so-called bottom-up approach. Both cases give valuable effective models since they permit to create a bigger landscape for a fundamental non-perturbative theory, that is unknown at present.    

%In this paper we will focus on the second approach used to describe the mass spectra for light scalar and vector mesons. 

One example of those nonperturbative phenomena is related to the dynamics of the lightest pseudoscalar mesons. A very useful Effective Field Theory approach  that describes it is given by the momentum-expansion formalism of Chiral Perturbation Theory (ChPT), where a $SU(N_{f})_{L}\otimes SU(N_{f})_{R}\rightarrow SU(N_{f})_{V}$-symmetric nonlinear sigma model (where $N_{f}=2,\,3$) written in terms of a meson multiplet is expanded up to a certain perturbative order; this procedure introduces a diagrammatic way to study scattering events between these particles in a particular range of energy \cite{Gasser:1983yg,Gasser:1984gg,Dobado:1997jx,Donoghuebook}. A phenomenological description is attained after fitting the parameters of the model (masses, decay constants and Low Energy Constants -LEC's-) to an adequate set of experimental data, e.g., phase shifts, scattering lenghts or imaginary parts of the associated amplitudes. 

The energy range mentioned above can be extended after unitarizing the partial waves of the scattering channels involved, thus including the respective resonances as poles in the complex plane \cite{Colangelo:2001df}. This method checks elastic unitarity in an approximate way (order by order in the expansion), although other approaches in the Momentum expansion allow to check exactly this feature, as happens with the Inverse Amplitude Method (IAM) \cite{Dobado:1989qm,Oller:1998hw}, in which pole positions are quite well described, specially those respecting pion-pion scalar and vector channels. 

Some of these resonances are properly analyzed as vector and tensor mesons since their structure is easily fitted as a Breit-Wigner distribution due to its $q\overline{q}$-like compositeness. However, quite the opposite happens with light scalar resonances $(I=0,\,J^{PC}=0^{++})$ produced below an energy close to 2 GeV since they are not easily characterized as $q\overline{q}$ mesons due to the large decay widths of some of these particles, as happens with the $f_{0}$ multiplet \cite{Olive:2016xmw}. This lies in the model-dependent descriptions of the nature of these particles, along with the inappropriate values for their masses and widths, as happens with the $f_{0}(500)$ when considering it as a $q\overline{q}$ meson in a $N_{f}=2$ linear sigma model \cite{Parganlija:2010fz}; nevertheless, recent approaches give a huge insight of the most likely nontrivial quark composition of this particle \cite{Pelaez:2015qba}. 

In what respects to pole positions, a proper model-independent description of the $f_{0}(500)$ and $f_{0}(980)$ resonance parameters is achieved by using an adequate set of dispersion relations with minimal uncertainty \cite{GarciaMartin:2011jx,GarciaMartin:2011cn}, thus theoretically minimizing the errors both in their masses and decay widths. Pole positions for the $f_{0}(500),\,f_{0}(980)\text{ and }f_{0}(1400/1370)$ can also be obtained through scattering matrix approaches, with the results depending on the way the couplings between scattering channels are taken into account \cite{Kaminski:1998ns}.  

Quark composition of resonances like $f_{0}(980),\,f_{0}(1370),\,f_{0}(1500)\text{ and }f_{0}(1710)$ can be studied via analysis of decay widths of $B$ mesons; in order to achieve this, these resonances have to be parametrized as superpositions of $u,d,s$ quarkonium states and a scalar glueball so that a perturbative QCD-effective Hamiltonian is to be built up, using their masses as the input parameters of the model (in this case, $f_{0}(980),\,f_{0}(1370)\text{ and }f_{0}(1710)$ are predominantly quarkonia) \cite{Wang:2006ria}. The $f_{0}(1500)$ is usually described as a glueball state since it does not decay into two photons (unlike what happens with the $f_{0}(1370)$) \cite{Branz:2009cv} and its mass coincides with lattice simulations \cite{Chen:2005mg}.  

Chiral effective models with a scalar glueball state are also used to study both compositeness and masses of scalar resonances where $m>\text{1.2 GeV}$ by considering experimental inputs such as the masses and composition of the scalar and pseudoscalar meson multiplets \cite{Giacosa:2005zt,Giacosa:2005qr}, with the results for quarkonia-glueball compositeness mixing for $f_{0}(1500)\text{ and }f_{0}(1710)$ depending on whether a glueball decay is or is not considered. In both cases, the theoretical masses of the resonances are quite close to the experimental values. Similar results for this mixing are obtained when considering lattice masses for quarkonia and glueball as input parameters \cite{Cheng:2006hu}.  

Mass generation for scalar resonances can be analyzed using a linear sigma model with two quark flavors including axial-vector mesons, a  glueball degree of freedom, and two parameters that explicitly break chiral and dilation symmetries, associated respectively to the $f_{0}(1370)$ and the $f_{0}(1500)$ (referred to as a dilaton field) \cite{Janowski:2011gt}. The results obtained after taking experimental inputs for quarkonia and glueball masses and widths come along with the compositeness mixing for these particles, hence giving that the $f_{0}(1500)$ is mostly a glueball state. For this case, the $f_{0}(1370)$ has a theoretical mass less than the lower experimental bound. A better result is obtained if the $f_{0}(1710)$ is considered as a glueball, however, this is discarded since the predicted value for its four pion decay width is large (something that has not been observed). If three quark flavors are to be taken \cite{Janowski:2014ppa},then the $f_{0}(1500)$ is considered as a heavy strange quarkonium, whereas the $f_{0}(1710)$ is largely composed by a glueball state. In this case, the glueball is coupled to meson states and mixed with two quarkonia states. After taking proper experimental inputs, the masses and widths of these three scalar particles are predicted within less than $10\%$ and $5\%$ of uncertainty, respectively. These results, along with the quarkonia-glueball mixing, are independent of the fit considered.  

When dealing with top-down models, an interesting approach considers a Dp/Dq deformed system that introduces a Hard Wall \cite{Wang:2009wx} or a Soft Wall \cite{Huang:2007fv} by choosing an specific dilaton profile, since dilatons are solutions arising from type IIB SUGRA background. In these sort of models, light scalar and vector mesons, and also glueballs, are well described when compared with lattice data \cite{Liu:2008ee}.  

In the case of bottom-up approaches, the most successful ones describing nonperturbative phenomena are the so-called AdS/QCD models, such as the hard wall (HW) \cite{BoschiFilho:2002ta} or the soft wall (SW) models \cite{Karch:2006pv} that were able to describe mass spectra, electromagnetic form factors, some decay constants, and other mesonic properties. The main idea behind these models is to break the conformal invariance in AdS by placing a cutoff, thus introducing an energy scale. When the cut-off is a D-brane, the model is called hard wall and when a quadratic dilaton is used instead, the soft wall model is obtained.

Results in the soft wall model show that masses grow linearly with the excitation number, which gives a Regge Trajectory. This mass spectrum appears due to the confining potential created by the quadratic dilaton profile \cite{Afonin:2012jn}.  When dealing with the hard wall model, the masses are given by the zeroes of Bessel functions generated by the Dirichlet boundary conditions imposed at the wall/brane, yielding non-linear trajectories \cite{Erdmenger:2007cm}.  Light vector mesons masses are described in \cite{Karch:2006pv} and scalar light mesons were described in \cite{Colangelo:2008us} in the soft wall model framework. These descriptions are not so good since they do not fit well the particle mass spectra, although mesons are organized in Regge trajectories \cite{Karch:2006pv,Colangelo:2008us}. 

Other soft wall approaches that consider scalar fields with variable masses (along with chiral symmetry effects) reproduce remarkable theoretical predictions for the light scalar sector when parameters such as quark masses and chiral condensates are introduced; however, in order to obtain these results, nonphysical values have to be taken into account for these set of parameters \cite{Vega:2010ne}. This issue is properly solved when a scalar potential is introduced \cite{Vega:2011tg}. Both of these results, besides reproducing quite well the light vector sector, consider the $f_{0}$ multiplet belonging to a Regge trajectory. 

Recently, a new approach was developed in \cite{Braga:2015jca}, where the usual soft wall model is upgraded by including an extra UV cutoff given by a D-brane. This extra brane will work as the boundary were the particles live and also will fix, altogether with the dilatonic energy scale, the mass and decay constant spectra of the particles. The application of this idea gives good results describing the first four vector states of charmonium and also the first four of  bottomonion  with a total error near to 30\% for fitting eight quarkonium states with three parameters \cite{Braga:2015jca}. The extension to finite temperature of this model gives a complete holographic view of the melting processes of these heavy quarkonium states, with results in agreement with the observed phenomenology \cite{Braga:2015jcb}. 

This paper is organized as follows: we introduce the holographic bottom-up model in section \ref{sec2} to describe the light scalar and vector meson resonances as poles of 2-point function. We show the main results of the model in section \ref{sec3}, regarding scalar and vector mass spectra, along with their respective error percentages when compared with experimental data. Finally, we present our conclusions in section \ref{sec4}. 

\section{Holographic Model for Light Mesons}
\label{sec2}

In order to describe light mesons, we will consider the usual SW model action \cite{Karch:2006pv,Colangelo:2008us}

\begin{align}
I=&-\frac{1}{2\,g_S^2}\int{d^5x\,\sqrt{-g}\,\exp[-\Phi\left(z\right)]\,\left[\partial_n\,S\,\partial^n\,S+m_5^2\,S^2\right]}\notag \\
&-\frac{1}{4\,g_V^2}\,\int{d^5x\,\sqrt{-g}\,\exp[-\Phi\left(z\right)]F_{mn}\,F^{mn}}, \label{general-action}
\end{align}

where $S\left(z,x^\mu\right)$ is a massive scalar field dual to the scalar mesons and $F_{mn}=\partial_m\,A_n-\partial_n\,A_m$ is given in terms of the massless abelian gauge field $A_m\left(z,x^\mu\right)$.

The bulk mass fixes the conformal dimension $\Delta$ of the $p-$form QCD operator $\mathcal{O}_s$ dual to  the $S$ field as $m_5^2\,R^2=\left(\Delta-p\right)\left(\Delta+p-4\right)$. In the simplest case, the scalar operator has the form $\mathcal{O}_s=\bar{q}\left(x\right)\,q\left(x\right)$ with dimension 3, where $q$ is any light quark. Thus, we can fix $\Delta=3$ and $p=0$ such that $m_5^2\,R^2=-3$ \cite{Colangelo:2008us}.

The geometric background is given by the sliced AdS Poincare patch \cite{Braga:2015jca,Braga:2015jcb}

\begin{equation}\label{geometry}
dS^2=\Theta\left(z-z_0\right)\,\frac{R^2}{z^2}\left[dz^2+\eta_{\mu\nu}\,dx^\mu\,dx^\nu\right],
\end{equation}

with $\Theta\left(z\right)$ the Heaviside step function that gives the UV D-brane (D-Wall) locus. The Minkowski metric has the signature $(-, +, +, +)$.

This particular choice of boundary for AdS breaks explicitly the conformal invariance by introducing an energy scale $z_0$, which can be associated to the nature of the strong interaction inside the meson \cite{Braga:2015jca}.  Such behavior is expected since when we recover the conformal boundary by setting $z_0\rightarrow 0$, the mass spectrum is given by an usual Regge trajectory defined by the form of the dilaton profile, i.e., $M_n^2 = c\left(n+s+1\right)$ \cite{Vega:2008te}. In this case, such profile corresponds to $\Phi\left(z\right)=\kappa^2z^2$, which is static as in the regular soft wall model. 

The constants $g_S$ and $g_V$ fix the units of the action in terms of the number of colors $N_c$ as usual. We shall not calculate decay constants nor form factors; those are quantities we are not interested in. The proper value for these couplings is read from the large 4-momentum expansion of the 2-point function in the QCD side compared to the same kind of expansion in the gravity side \cite{Karch:2006pv,Colangelo:2008us}. 

Following the ideas exposed in \cite{Braga:2015jca}, we will define the mass spectrum of light scalar and vector mesons as functions of two energy scales, namely, the D-wall locus $z_0$ and the dilaton constant $\kappa$. 

\subsection{Light Vector Mesons}
\label{subsec-vector}

We begin our analysis with the light vector meson action given by 

\begin{equation}
I_{V}=-\frac{1}{4\,g_V^2}\,\int{d^5x\,\sqrt{-g}\,\exp[-\Phi\left(z\right)]F_{mn}\,F^{mn}},
\label{Vector-action}
\end{equation}

according to (\ref{general-action}). After considering small variations in the $A_{\mu}$ field and imposing the gauge condition $A_z=0$, we obtain  the equation of motion for the space-time components as 

\begin{equation}
\partial_{z}\left[\frac{\exp(-\kappa^{2}z^{2})}{z}\,\partial_{z}A^{\mu}\right]+\frac{\exp(-\kappa^{2}z^{2})}{z}\,\eta^{\rho\sigma}\,\partial_{\rho}\partial_{\sigma}A^{\mu}=0.
\label{vectoreom}
\end{equation}

Equation (\ref{vectoreom}) allows us to obtain a boundary action from (\ref{Vector-action}) for the vector fields that reads

\begin{equation}
I_{\text{V On-Shell}}=-\frac{R}{2g_{V}^{2}}\int{d^{5}x\left\{\partial_{z}\left[\frac{\exp(-\kappa^{2}z^{2})}{z}\,A_{n}\,\partial_{z}A^{n}\right]\right\}}.
\label{Vector-osaction}
\end{equation}

In the latter equation, we have used again the gauge condition $A_{z}=0$. According to the Minkowskian prescription, this boundary action (\ref{Vector-osaction}) gives the 2-point function, and its poles define the mass spectrum.  From this same equation, we infer that the boundary term (i.e., taking $z=z_{0}$) is such that

\begin{equation}
I_{\text{V On-Shell}}^{\text{Boundary}}=-\frac{R}{2g_{V}^{2}}\int{d^{4}x\left. \frac{\exp(-\kappa^{2}z^{2})}{z}A_{\mu}\,\partial_{z}\,A^{\mu}\right|_{z_{0}}}.
\label{Vec-osbdaction}
\end{equation}

Two-point functions are easily obtained after solving the equation of motion (\ref{vectoreom}) by introducing Fourier transformed vector fields 

\begin{equation}
A^{\mu}(z,x^{\mu})=\frac{1}{(2\pi)^{4}}\int{d^{4}q\,\exp(-iq_{\mu}x^{\mu})\,\,v_{\mu}(z,q)},
\label{vector-ft}
\end{equation}

where we write $v_{\mu}(z,q)$ as a function of the source term $v_{\mu}^{0}(q)$ and the Bulk-to-Boundary propagator $V(z,q)$ as follows:

\begin{equation}
v_{\mu}(z,q)=v_{\mu}^{0}\left(q\right)\,V(z,q).
\label{Fourier-vector}
\end{equation}

Therefore, and reminding that $\eta^{\rho\sigma}\partial_{\sigma}\partial_{\rho}=-\square=q^{2}$, we obtain that $V(z,q)$ holds with the following:

\begin{equation}
\partial_{z}\left[\frac{\exp(-\kappa^{2}z^{2})}{z}\,\partial_{z}V(z,q)\right]+\frac{q^{2}}{z}\,\exp(-\kappa^{2}z^{2})V(z,q)=0.
\label{eom-BtoBp}
\end{equation}

The regular solution of (\ref{eom-BtoBp}) reads

\begin{equation}
V(z,q)=c_{1}\,\kappa^{2}\,z^{2}\,\text{}_{\,1}F_{1}\left(1-\frac{q^{2}}{4\kappa^{2}},2,\kappa^{2}z^{2}\right),
\label{vec-regsol}
\end{equation}

where $\text{}_{\,1}F_{1}(1-q^{2}/4\kappa^{2},2,\kappa^{2}z^{2})$ is the Kummer confluent hypergeometric function and $c_{1}$ is a normalization constant. Hence, we deduce from the On-Shell boundary action 

\begin{equation}
I_{\text{V On-Shell}}^{\text{Boundary}}=-\frac{R}{2g_{V}^{2}}\int\frac{d^{4}q}{(2\pi)^{4}}\,v_{\mu}^{0}(q)v^{\mu\, 0}(-q)\left. \frac{\exp(-\kappa^{2}z^{2})}{z}V(z,q)\,\partial_{z}V(z,-q)\right|_{z_{0}}
\label{Vec-osbdaction2}
\end{equation}

the following vector two-point function $G^{\mu\nu}(q^{2})$.

\begin{align}
&G^{\mu\nu}(q^{2})=\eta^{\mu\nu}\,\Pi(q^{2}), \label{vec-twopoint} \\
&\Pi(q^{2})=-\frac{R}{g_{V}^{2}}\left. \left[\frac{\exp(-\kappa^{2}z^{2})}{z}\,V(z,q)\,\partial_{z}\,V(z,-q)\right]\right|_{z_{0}}. \label{pi-factor}
\end{align}

After normalizing (\ref{vec-regsol}) such that $V(z_{0})=1$, we finally obtain that $\Pi(q^{2})$ reads

\begin{equation}
\Pi(q^{2})=-\frac{R\,\exp(-\kappa^{2}z_{0}^{2})}{g_{V}^{2}z_{0}^{2}}\left[\frac{2}{z_{0}}+\kappa^{2}z_{0}\left(1-\frac{q^{2}}{4\kappa^{2}}\right)\frac{_{1}F_{1}\left(2-\frac{q^{2}}{4\kappa^{2}},3,\kappa^{2}z_{0}^{2}\right)}{_{1}F_{1}\left(1-\frac{q^{2}}{4\kappa^{2}},2,\kappa^{2}z_{0}^{2}\right)}\right].
\label{pi-factor2}
\end{equation} 

The poles of the 2-point function (\ref{pi-factor2}) can be read from the roots  of the hypergeometric confluent function in the denominator

\begin{equation}
\text{}_{\,1}F_{1}\left(1-\chi_n,2,\kappa^2\,z_0^2\right)=0,
\end{equation}

with $\chi_{n}=q_n^2/4\,\kappa^2$ the root spectrum  and  $q_n^2=M_n^2$ the physical masses. Thus, the mass spectrum for light vector mesons is given by

\begin{equation}\label{vector-mass-spectrum}
M^2_{n,\text{V}}=4\,\kappa^2\chi_n\left(z_0,\kappa\right).
\end{equation}

The result above assures that the mass spectrum (\ref{vector-mass-spectrum}) is given by a non-linear Regge trajectory defined by the parameters $z_0$ and $\kappa$. In general, the roots of the hypergeometric confluent function increase with $n$ \cite[Sec. 13.9]{NIST:DLMF}, so the masses increase with the excitation number, as we expected. The results for the light vector masses are showed in Table \ref{tab:vecmes}.

\subsection{Light Scalar Mesons}

We see that the scalar case follows a similar procedure that the vector fields showed in \ref{subsec-vector}. Thus, we define from (\ref{general-action}) the scalar action as

\begin{equation}
I_{S}=-\frac{1}{2g_{S}^{2}}\int{d^{5}x\,\sqrt{-g}\,\exp[-\Phi\left(z\right)]\,\left[\partial_n\,S\,\partial^n\,S+m_5^2\,S^2\right]},
\label{scalar-action}
\end{equation} 

whose associated equation of motion, after taking  small variations in $S$, the gauge condition $A_z=0$ and replacing the definition of the conformal dimension in terms of $m_{S}$, is given by

\begin{equation}
\partial_{z}\left[\frac{\exp(-\kappa^{2}z^{2})}{z}\partial_{z}S\right]-\frac{\exp(-\kappa^{2}z_{2})}{z^{3}}\,\square S+\frac{3\exp(-\kappa^{2}z^{2})}{z^{5}}S=0,
\label{scalar-eom}
\end{equation}

where $\square=-\eta^{\mu\nu}\partial_{\mu}\partial_{\nu}$. We obtain the solution of (\ref{scalar-eom}) by considering the Fourier transform of the scalar field as

\begin{align}
&S(x_{\mu},z)=\frac{1}{(2\pi)^{4}}\int{\exp(-ix_{\mu}q^{\mu})S(z,q)}, \label{scalar-fourier} \\
&S(z,q)=S^{0}(q)\overline{v}(z,q) \label{scalar-source}.
\end{align}

In this case, the Bulk-to-Boundary propagator is labeled as $\overline{v}(z,q)$, whilst the scalar source term is given by $S^{0}(q)$. Hence, (\ref{scalar-eom}) changes into

\begin{equation}
\partial_{z}\left[\frac{\exp(-\kappa^{2}z^{2})}{z^{3}}\partial_{z}\overline{v}(z,q\right]+\frac{\exp(-\kappa^{2}z^{2})}{z^{3}}q^{2}\overline{v}(z,q)+\frac{3\exp(-\kappa^{2}z^{2})}{z^{5}}\overline{v}(z,q)=0,
\label{scalar-eom-vz}
\end{equation}

whose regular solution is given in terms of the Kummer confluent hypergeometric function as follows:

\begin{equation}
\overline{v}(z,q)=\overline{c}_{1}\,\kappa^{3}z^{3}\,\text{}_{1}F_{1}\left(\frac{3}{2}-\frac{q^{2}}{4\kappa^{2}},2,\kappa^{2}z^{2}\right).
\label{scalar-propbtb}
\end{equation}

As expected, our solution depends on a normalization constant $\overline{c}_{1}$. Before showing the normalized solution of the Bulk-to-Boundary propagator, we deduce from (\ref{scalar-action}) that the On-Shell Boundary action reads

\begin{equation}
I_{\text{S On-Shell}}^{\text{Boundary}}=\frac{R^{3}}{g_{S}^{2}}\int{d^{4}q\left. \frac{\exp(-\kappa^{2}z^{2})}{z^{3}}\,S^{0}(q)\,S^{0}(-q)\,\overline{v}(z,q)\,\partial_{z}\overline{v}(z,-q)\right|_{z_{0}}}.
\label{scalar-actionbound}
\end{equation}

Hence, the scalar two-point function $\Pi_{S}(q^{2})$ is such that

\begin{equation}
\Pi_{S}(q^{2})=-\frac{R^{3}}{g_{S}^{2}}\left. \frac{\exp(-\kappa^{2}z^{2})}{z^{3}}\,\overline{v}(z,q)\,\partial_{z}\,\overline{v}(z,-q)\right|_{z_{0}}.
\label{scalar-twop}
\end{equation}

Our solution for (\ref{scalar-twop}), written in terms of a normalized $\overline{v}(z,q)$ function, is given by

\begin{equation}
\Pi_{S}(q^{2})=-\frac{R^{3}}{g_{S}^{2}}\frac{\exp(-\kappa^{2}z_{0}^{2})}{z_{0}^{\,3}}\left[\frac{3}{z_{0}}+\kappa^{2}z_{0}\left(\frac{3}{2}-\frac{q^{2}}{4\kappa^{2}}\right)\frac{_{1}F_{1}\left(\frac{5}{2}-\frac{q^{2}}{4\kappa^{2}},3,\kappa^{2}z_{0}^{2}\right)}{_{1}F_{1}\left(\frac{3}{2}-\frac{q^{2}}{4\kappa^{2}},2,\kappa^{2}z_{0}^{2}\right)}\right].
\label{scalar-twop2}
\end{equation}

As in the vector case, we obtain the pole expansion from the roots of the denominator in (\ref{scalar-twop2})

\begin{equation}
_{1}F_{1}\left(\frac{3}{2}-\overline{\chi}_n,2,\kappa^2\,z_0^2\right)=0,
\end{equation}

with $\overline{\chi}_n=q_n^2/4\kappa$. Therefore, the mass spectrum is given by 

\begin{equation} \label{scalar-mass-spectrum}
M_{n, \text{S}}^2=4\,\kappa^2\,\overline{\chi}_n\left(z_0,\kappa\right).
\end{equation}

Notice that (\ref{scalar-mass-spectrum}) is also non-linear and defined by the increasing $\overline{\chi}_n$ and the parameters $\kappa$ and $z_0$.  The results for these mesons are showed in Table \ref{tab:scalmes}.

\section{Results}
\label{sec3}

\begin{table}
\centering
\begin{tabular}{ccccccc}   \hline \hline
	$\rho$ & & $M_{\text{th}}$ (GeV)& & $M_{\text{exp}}$ (GeV) & & $\% M$  \\ \hline
	$\rho(775)$ & & 0.975  & & 0.775 & & 20.53  \\ 
	$\rho(1450)$  & & 1.455 & & 1.465  & & 0.66 \\
	$\rho(1570)$ & & 1.652 & & 1.570 & & 4.96 \\
	$\rho(1700)$ & & 1.829 & & 1.720 & & 5.97 \\ 
	$\rho(1900)$ & & 1.992 & & 1.909 & & 4.15 \\
	$\rho(2150)$ & & 2.142 & & 2.153 & & 0.50   \\ \hline \hline
\end{tabular}
\caption{Mass spectrum for $\rho$ vector mesons with $\kappa=0.45$ GeV and $z_0=5$ GeV$^{-1}$. Experimental values are obtained from \cite{Olive:2016xmw}. }
\label{tab:vecmes}
\end{table}

The respective spectra for vector and scalar resonances is generated after finding the associated poles of the two-point functions (\ref{pi-factor2}) and (\ref{scalar-twop2}). In order to obtain them, we only need to fix two parameters: the boundary radius $z_{0}$ and the dilaton slope $\kappa$. Following \cite{Braga:2015jca}, we will fix $\kappa$  as flavor independent, so we will use the same $\kappa$ for scalar and vector mesons since they are made of up and down quarks, which in the chiral limit have the same mass.   The $z_0$ parameter is defined as a quantity related to the nature of the strong interactions inside the mesons. Thus, we could use the same value reported in \cite{Braga:2015jca}, but due to the color screening it is expected that the $z_0$ parameter would be different for light and heavy quarks. 

In this case, we have that the best values that fits the experimental masses \cite{Olive:2016xmw} correspond to 

\begin{align}
&z_{0}=\text{5 GeV}^{-1}, \\
&\kappa=\text{0.45 GeV}.
\end{align}

In Table \ref{tab:vecmes}, we present the theoretical values calculated with the model proposed in \cite{Braga:2015jca}, along with the experimental masses and the corresponding uncertainties for the $\rho$ vector meson trajectory. It is interesting to notice that the spectrum is not linear, as in the case of the regular soft wall model \cite{Karch:2006pv}. 

We show in Table \ref{tab:scalmes} the results for the $f_0$ trajectory. Again, the spectrum is nonlinear. Notice that the $n=1$ state is not associated to the $f_0\left(500\right)$ state. In this model, it is not possible to fit this resonance into the trajectory (\ref{scalar-mass-spectrum}) with any parameter choice. Thus, since we have related $\kappa$ and $z_0$ with the color structure inside mesons, we can conclude that, holographically, the $f_0\left(500\right)$ resonance is not a $q\bar{q}$ state. This is in agreement with theoretical phenomenology \cite{Pelaez:2015qba}.   

Following \cite{Braga:2015jca}, we can test the predictability of the model developed here with the RMS error for estimating $N$ parameters using $N_p$ parameters as

\begin{equation}
\delta_\text{RMS}=\sqrt{\frac{1}{N-N_p}\,\sum_i^{N}\left(\frac{\delta\,\mathcal{O}_i}{\mathcal{O}_i}\right)^2},
\end{equation}

where $\mathcal{O}_i$ is the experimental mean value of a given observable and $\delta\,\mathcal{O}_i$ is the absolute uncertainty given by the model. In our case, we fit up to 14 resonant states with  two parameters, thus obtaining an RMS error $\delta_{\text{RMS}}$

\begin{equation}
\delta_{\text{RMS}}=\text{7.64}\%.
\label{deltarms}
\end{equation}

\begin{table}
\centering
\begin{tabular}{ccccccc}   \hline \hline
	$f_{0}$ & & $M_{\text{th}}$ (MeV)& & $M_{\text{exp}}$ (MeV) & & $\% M$  \\ \hline
	$f_{0}(980)$ & & 1.070  & & 0.99 & & 7.46  \\ 
	$f_{0}(1370)$  & & 1.284 & & 1.370  & &  5.11 \\
	$f_{0}(1500)$ & & 1.487 & & 1.504 & & 1.13 \\
	$f_{0}(1710)$ & & 1.674 & & 1.723 & & 2.93 \\ 
	$f_{0}(2020)$ & & 1.846 & & 1.992 & & 7.94 \\
	$f_{0}(2100)$ & & 2.153 & & 2.101 & & 2.39   \\ 
	$f_{0}(2200)$ & & 2.292 & & 2.189 & & 4.49   \\  
	$f_{0}(2330)$ & & 2.424 & & 2.314 & & 4.52   \\  \hline \hline
\end{tabular}
\caption{Mass spectrum for $f_{0}$ scalar resonances with $\kappa=0.45$ GeV and $z_0=5.0$ GeV$^{-1}$. Experimental values for the masses are read from \cite{Olive:2016xmw}.}
\label{tab:scalmes}
\end{table}

As it can be seen from Tables \ref{tab:vecmes} and \ref{tab:scalmes}, the resonances we obtain are not degenerate, as expected from the usual Regge theory. We attain this after carefully choosing the pole positions of the two-point functions (\ref{pi-factor2}) and (\ref{scalar-twop2}) according to their $q^{2}$-dependence.  

We also want to point out that the approach considered here minimizes the amount of parameters to be taken into account since the model (both in the scalar and vector sector) does not deal directly with a certain meson internal structure, as showed in \cite{Vega:2010ne,Vega:2011tg} (All this information is summarized in the choosing of the $(\kappa,z_{0})$ parameter space). Pions and axial states are not reproduced since we do not take into account chiral symmetry breaking effects.

\section{Conclusions}
\label{sec4}

The model we considered does not deal directly with the composition of the scalar mesons, as in the case of the $f_{0}(1500)$ and $f_{0}(1710)$, which are glueball candidates. This was not necessary since the poles only depend on the model parameters $\kappa$ and $z_{0}$. Besides, the errors we obtained are within the phenomenological bounds given in \cite{Janowski:2014ppa}. We also obtained a remarkable result for the $f_{0}(980)$ mass, a possible non-$q\overline{q}$ state. However, the $f_{0}(500)$ was not possible to fit; this means that the model needs to be somehow extended to describe these sort of scalar particles. On the other hand, light vector mesons were well fitted, with the $\rho(770)$ state having the biggest error. As a matter of fact, and unlike what happened with the scalar multiplet, the ground state could be determined up to the higher error bound allowed by these sort of nonconformal models. We also remark that all of our results did not need to consider either experimental or lattice input parameters.

We showed here that these AdS/QCD approaches could reproduce light meson spectra after minimizing the amount of holographic and physical parameters; we attained this by analyzing the respective poles of the scalar and vector propagators in such a way that only the dilaton profile $\kappa$ and the D-wall locus $z_{0}$ are needed, thus avoiding the introduction of nonphysical quark masses and condensates, as in \cite{Vega:2010ne}. Furthermore, internal properties of mesons were also avoided here since quark masses and condensates-dependent confining potentials \cite{Vega:2011tg} were not directly treated. These parameters are, by some unknown form, related with the constituent quark mass and to the natureness of the strong interaction.

Despite having different values for $\kappa$ and $z_{0}$ for heavy \cite{Braga:2015jca} and light mesons, an universality class can be established for these sort of models. In fact, there is a huge  phenomelogical difference between heavy and light quarks due to the heavy quark symmetry: heavy quark systems are considered as non-relativistic, e.g., Schroedinger-like heavy quarkonium potentials. Besides, color screening effects in both systems are different since they strongly depend on the quark masses \cite{Shuryak:1988ck}.

As a future work, we want to study finite-temperature chiral symmetry restoration effects in these sort of models \cite{Braga:2015jcb} after properly introducing pseudoscalar and axial particles. Our objective is checking if these holographic approaches properly describe phase transitions, as happens with large-$N$ nonlinear sigma models \cite{Cortes:2015emo,Cortes:2016ecy}.

\section*{Acknowledgments}
 
We want to thank Facultad de Ciencias and Vicerrector\'{\i}a de Investigaci\'on of Universidad de los Andes for financial support.


\begin{thebibliography}{99}

%\cite{Maldacena:1997re}
\bibitem{Maldacena:1997re} 
  J.~M.~Maldacena,
  %``The Large N limit of superconformal field theories and supergravity,''
  Int.\ J.\ Theor.\ Phys.\  {\bf 38}, 1113 (1999)
  [Adv.\ Theor.\ Math.\ Phys.\  {\bf 2}, 231 (1998)].
  %doi:10.1023/A:1026654312961
  %[hep-th/9711200].
  %%CITATION = doi:10.1023/A:1026654312961;%%
  %12782 citations counted in INSPIRE as of 31 May 2017
	

%\cite{Ramallo:2013bua}	
\bibitem{Ramallo:2013bua} 
  A.~V.~Ramallo,
  ``Introduction to the AdS/CFT correspondence,''
  Springer Proc.\ Phys.\  {\bf 161}, 411 (2015)
  %doi:10.1007/978-3-319-12238-0_10
  [arXiv:1310.4319 [hep-th]].
  %%CITATION = doi:10.1007/978-3-319-12238-0_10;%%
  %41 citations counted in INSPIRE as of 31 May 2017

%\cite{Gasser:1983yg}
\bibitem{Gasser:1983yg} 
  J.~Gasser and H.~Leutwyler,
  %``Chiral Perturbation Theory to One Loop,''
  Annals Phys.\  {\bf 158}, 142 (1984).
  %doi:10.1016/0003-4916(84)90242-2
  %%CITATION = doi:10.1016/0003-4916(84)90242-2;%%
  %3621 citations counted in INSPIRE as of 26 Sep 2016
  
 
 %\cite{Gasser:1984gg}
\bibitem{Gasser:1984gg} 
  J.~Gasser and H.~Leutwyler,
  %``Chiral Perturbation Theory: Expansions in the Mass of the Strange Quark,''
  Nucl.\ Phys.\ B {\bf 250}, 465 (1985).
  %doi:10.1016/0550-3213(85)90492-4
  %%CITATION = doi:10.1016/0550-3213(85)90492-4;%%
  %3449 citations counted in INSPIRE as of 12 Oct 2016
	
	
%\cite{Dobado:1997jx}
\bibitem{Dobado:1997jx} 
  A.~Dobado, A.~Gomez-Nicola, A.~L.~Maroto and J.~R.~Pelaez,
  ``Effective lagrangians for the standard model,''
  N.Y., Springer-Verlag, 1997. (Texts and Monographs in Physics)
  %17 citations counted in INSPIRE as of 01 May 2017
	
\bibitem{Donoghuebook} 
  J. F. Donoghue, E. Golowich and B. R. Holstein,
  ``Dynamics of the Standard Model,''
	Cambridge University Press, Australia, 1992.

  
 %\cite{Colangelo:2001df}
\bibitem{Colangelo:2001df} 
  G.~Colangelo, J.~Gasser and H.~Leutwyler,
  %``$\pi \pi$ scattering,''
  Nucl.\ Phys.\ B {\bf 603}, 125 (2001).
  %doi:10.1016/S0550-3213(01)00147-X
  %[hep-ph/0103088].
  %%CITATION = doi:10.1016/S0550-3213(01)00147-X;%%
  %755 citations counted in INSPIRE as of 18 Oct 2016
 

%\cite{Dobado:1989qm}
\bibitem{Dobado:1989qm} 
  A.~Dobado, M.~J.~Herrero and T.~N.~Truong,
  %``Unitarized Chiral Perturbation Theory for Elastic Pion-Pion Scattering,''
  Phys.\ Lett.\ B {\bf 235}, 134 (1990).
  %doi:10.1016/0370-2693(90)90109-J
  %%CITATION = doi:10.1016/0370-2693(90)90109-J;%%
  %278 citations counted in INSPIRE as of 19 Oct 2016

 
 
 %\cite{Oller:1998hw}
\bibitem{Oller:1998hw} 
  J.~A.~Oller, E.~Oset and J.~R.~Pelaez,
  %``Meson meson interaction in a nonperturbative chiral approach,''
  Phys.\ Rev.\ D {\bf 59}, 074001 (1999)
  Erratum: [Phys.\ Rev.\ D {\bf 60}, 099906 (1999)]
  Erratum: [Phys.\ Rev.\ D {\bf 75}, 099903 (2007)].
  %doi:10.1103/PhysRevD.59.074001, 10.1103/PhysRevD.60.099906, 10.1103/PhysRevD.75.099903
  %[hep-ph/9804209].
  %%CITATION = doi:10.1103/PhysRevD.59.074001, 10.1103/PhysRevD.60.099906, 10.1103/PhysRevD.75.099903;%%
  %560 citations counted in INSPIRE as of 19 Oct 2016
	

%\cite{Olive:2016xmw}
\bibitem{Olive:2016xmw} 
  C.~Patrignani {\it et al.} [Particle Data Group],
  %``Review of Particle Physics,''
  Chin.\ Phys.\ C {\bf 40}, no. 10, 100001 (2016).
  %doi:10.1088/1674-1137/40/10/100001
  %%CITATION = doi:10.1088/1674-1137/40/10/100001;%%
  %758 citations counted in INSPIRE as of 01 May 2017
	
	
%\cite{Parganlija:2010fz}
\bibitem{Parganlija:2010fz} 
  D.~Parganlija, F.~Giacosa and D.~H.~Rischke,
  %``Vacuum Properties of Mesons in a Linear Sigma Model with Vector Mesons and Global Chiral Invariance,''
  Phys.\ Rev.\ D {\bf 82}, 054024 (2010).
  %doi:10.1103/PhysRevD.82.054024
  %[arXiv:1003.4934 [hep-ph]].
  %%CITATION = doi:10.1103/PhysRevD.82.054024;%%
  %103 citations counted in INSPIRE as of 02 May 2017
 

%\cite{Pelaez:2015qba}
\bibitem{Pelaez:2015qba} 
  J.~R.~Pelaez,
  %``From controversy to precision on the sigma meson: a review on the status of the non-ordinary $f_0(500)$ resonance,''
  Phys.\ Rept.\  {\bf 658}, 1 (2016).
  %doi:10.1016/j.physrep.2016.09.001
  %[arXiv:1510.00653 [hep-ph]].
  %%CITATION = doi:10.1016/j.physrep.2016.09.001;%%
  %84 citations counted in INSPIRE as of 01 May 2017
	
	
%\cite{GarciaMartin:2011jx}
\bibitem{GarciaMartin:2011jx} 
  R.~Garcia-Martin, R.~Kaminski, J.~R.~Pelaez and J.~Ruiz de Elvira,
  %``Precise determination of the f0(600) and f0(980) pole parameters from a dispersive data analysis,''
  Phys.\ Rev.\ Lett.\  {\bf 107}, 072001 (2011).
  %doi:10.1103/PhysRevLett.107.072001
  %[arXiv:1107.1635 [hep-ph]].
  %%CITATION = doi:10.1103/PhysRevLett.107.072001;%%
  %146 citations counted in INSPIRE as of 01 May 2017
	
	
%\cite{GarciaMartin:2011cn}
\bibitem{GarciaMartin:2011cn} 
  R.~Garcia-Martin, R.~Kaminski, J.~R.~Pelaez, J.~Ruiz de Elvira and F.~J.~Yndurain,
  %``The Pion-pion scattering amplitude. IV: Improved analysis with once subtracted Roy-like equations up to 1100 MeV,''
  Phys.\ Rev.\ D {\bf 83}, 074004 (2011).
  %doi:10.1103/PhysRevD.83.074004
  %[arXiv:1102.2183 [hep-ph]].
  %%CITATION = doi:10.1103/PhysRevD.83.074004;%%
  %201 citations counted in INSPIRE as of 01 May 2017
	
	
%\cite{Kaminski:1998ns}
\bibitem{Kaminski:1998ns} 
  R.~Kaminski, L.~Lesniak and B.~Loiseau,
  %``Scalar mesons and multichannel amplitudes,''
  Eur.\ Phys.\ J.\ C {\bf 9}, 141 (1999).
  %doi:10.1007/s100530050414, 10.1007/s100529900023
  %[hep-ph/9810386].
  %%CITATION = doi:10.1007/s100530050414, 10.1007/s100529900023;%%
  %84 citations counted in INSPIRE as of 01 May 2017
	
	
%\cite{Wang:2006ria}
\bibitem{Wang:2006ria} 
  W.~Wang, Y.~L.~Shen, Y.~Li and C.~D.~Lu,
  %``Study of scalar mesons f0(980) and f0(1500) from B ---> f0(980) K and B ---> f0(1500) K Decays,''
  Phys.\ Rev.\ D {\bf 74}, 114010 (2006).
  %doi:10.1103/PhysRevD.74.114010
  %[hep-ph/0609082].
  %%CITATION = doi:10.1103/PhysRevD.74.114010;%%
  %33 citations counted in INSPIRE as of 02 May 2017
	
	
%\cite{Branz:2009cv}
\bibitem{Branz:2009cv} 
  T.~Branz, L.~S.~Geng and E.~Oset,
  %``Two-photon and one photon-one vector meson decay widths of the f(0)(1370), f(2)(1270), f(0)(1710), f-prime(2)(1525), and K*(2)(1430),''
  Phys.\ Rev.\ D {\bf 81}, 054037 (2010).
  %doi:10.1103/PhysRevD.81.054037
  %[arXiv:0911.0206 [hep-ph]].
  %%CITATION = doi:10.1103/PhysRevD.81.054037;%%
  %29 citations counted in INSPIRE as of 02 May 2017
	
	
%\cite{Chen:2005mg}
\bibitem{Chen:2005mg} 
  Y.~Chen {\it et al.},
  %``Glueball spectrum and matrix elements on anisotropic lattices,''
  Phys.\ Rev.\ D {\bf 73}, 014516 (2006).
  %doi:10.1103/PhysRevD.73.014516
  %[hep-lat/0510074].
  %%CITATION = doi:10.1103/PhysRevD.73.014516;%%
  %360 citations counted in INSPIRE as of 02 May 2017
	
	
%\cite{Giacosa:2005zt}
\bibitem{Giacosa:2005zt} 
  F.~Giacosa, T.~Gutsche, V.~E.~Lyubovitskij and A.~Faessler,
  %``Scalar nonet quarkonia and the scalar glueball: Mixing and decays in an effective chiral approach,''
  Phys.\ Rev.\ D {\bf 72}, 094006 (2005).
  %doi:10.1103/PhysRevD.72.094006
  %[hep-ph/0509247].
  %%CITATION = doi:10.1103/PhysRevD.72.094006;%%
  %123 citations counted in INSPIRE as of 02 May 2017
	
	
	
%\cite{Giacosa:2005qr}
\bibitem{Giacosa:2005qr} 
  F.~Giacosa, T.~Gutsche, V.~E.~Lyubovitskij and A.~Faessler,
  %``Scalar meson and glueball decays within a effective chiral approach,''
  Phys.\ Lett.\ B {\bf 622}, 277 (2005).
  %doi:10.1016/j.physletb.2005.07.016
  %[hep-ph/0504033].
  %%CITATION = doi:10.1016/j.physletb.2005.07.016;%%
  %91 citations counted in INSPIRE as of 02 May 2017
	
	
%\cite{Cheng:2006hu}
\bibitem{Cheng:2006hu} 
  H.~Y.~Cheng, C.~K.~Chua and K.~F.~Liu,
  %``Scalar glueball, scalar quarkonia, and their mixing,''
  Phys.\ Rev.\ D {\bf 74}, 094005 (2006).
  %doi:10.1103/PhysRevD.74.094005
  %[hep-ph/0607206].
  %%CITATION = doi:10.1103/PhysRevD.74.094005;%%
  %108 citations counted in INSPIRE as of 02 May 2017
	
	
%\cite{Janowski:2011gt}
\bibitem{Janowski:2011gt} 
  S.~Janowski, D.~Parganlija, F.~Giacosa and D.~H.~Rischke,
  %``The Glueball in a Chiral Linear Sigma Model with Vector Mesons,''
  Phys.\ Rev.\ D {\bf 84}, 054007 (2011).
  %doi:10.1103/PhysRevD.84.054007
  %[arXiv:1103.3238 [hep-ph]].
  %%CITATION = doi:10.1103/PhysRevD.84.054007;%%
  %75 citations counted in INSPIRE as of 02 May 2017
	
	
%\cite{Janowski:2014ppa}
\bibitem{Janowski:2014ppa} 
  S.~Janowski, F.~Giacosa and D.~H.~Rischke,
  %``Is f0(1710) a glueball?,''
  Phys.\ Rev.\ D {\bf 90}, no. 11, 114005 (2014).
  %doi:10.1103/PhysRevD.90.114005
  %[arXiv:1408.4921 [hep-ph]].
  %%CITATION = doi:10.1103/PhysRevD.90.114005;%%
  %58 citations counted in INSPIRE as of 02 May 2017
  
 %\cite{Wang:2009wx}
\bibitem{Wang:2009wx} 
  C.~Wang, S.~He, M.~Huang, Q.~S.~Yan and Y.~Yang,
  %``Scalar Mesons and glueballs in Dp-Dq hard-wall models,''
  Chin.\ Phys.\ C {\bf 34}, 319 (2010).
  %doi:10.1088/1674-1137/34/3/003
  %[arXiv:0902.0864 [hep-ph]].
  %%CITATION = doi:10.1088/1674-1137/34/3/003;%%
  %2 citations counted in INSPIRE as of 12 Jul 2017
  
 
 %\cite{Huang:2007fv}
\bibitem{Huang:2007fv} 
  S.~He, M.~Huang, Q.~S.~Yan and Y.~Yang,
  %``Confront Holographic QCD with Regge Trajectories,''
  Eur.\ Phys.\ J.\ C {\bf 66}, 187 (2010).
  %doi:10.1140/epjc/s10052-010-1239-0
  %[arXiv:0710.0988 [hep-ph]].
  %%CITATION = doi:10.1140/epjc/s10052-010-1239-0;%%
  %29 citations counted in INSPIRE as of 12 Jul 2017
  
  
 %\cite{Liu:2008ee}
\bibitem{Liu:2008ee} 
  K.~F.~Liu,
  %``Challenges of Lattice Calculation of Scalar Mesons,''
  AIP Conf.\ Proc.\  {\bf 1030}, 305 (2008)
  %doi:10.1063/1.2973517
  [arXiv:0805.3364 [hep-lat]].
  %%CITATION = doi:10.1063/1.2973517;%%
  %13 citations counted in INSPIRE as of 12 Jul 2017
 

%\cite{BoschiFilho:2002ta}
\bibitem{BoschiFilho:2002ta} 
  H.~Boschi-Filho and N.~R.~F.~Braga,
  %``QCD / string holographic mapping and glueball mass spectrum,''
  Eur.\ Phys.\ J.\ C {\bf 32}, 529 (2004).
  %doi:10.1140/epjc/s2003-01526-4
  %[hep-th/0209080].
  %%CITATION = doi:10.1140/epjc/s2003-01526-4;%%
  %125 citations counted in INSPIRE as of 20 Jun 2017 

%\cite{Karch:2006pv}
\bibitem{Karch:2006pv} 
  A.~Karch, E.~Katz, D.~T.~Son and M.~A.~Stephanov,
  %``Linear confinement and AdS/QCD,''
  Phys.\ Rev.\ D {\bf 74}, 015005 (2006).
  %doi:10.1103/PhysRevD.74.015005
  %[hep-ph/0602229].
  %%CITATION = doi:10.1103/PhysRevD.74.015005;%%
  %718 citations counted in INSPIRE as of 20 Jun 2017

%\cite{Afonin:2012jn}
\bibitem{Afonin:2012jn} 
  S.~S.~Afonin,
  %``Generalized Soft Wall Model,''
  Phys.\ Lett.\ B {\bf 719}, 399 (2013).
  %doi:10.1016/j.physletb.2013.01.055
  %[arXiv:1210.5210 [hep-ph]].
  %%CITATION = doi:10.1016/j.physletb.2013.01.055;%%
  %10 citations counted in INSPIRE as of 20 Jun 2017 
	

%\cite{Erdmenger:2007cm}
\bibitem{Erdmenger:2007cm} 
  J.~Erdmenger, N.~Evans, I.~Kirsch and E.~Threlfall,
  %``Mesons in Gauge/Gravity Duals - A Review,''
  Eur.\ Phys.\ J.\ A {\bf 35}, 81 (2008).
  %doi:10.1140/epja/i2007-10540-1
  %[arXiv:0711.4467 [hep-th]].
  %%CITATION = doi:10.1140/epja/i2007-10540-1;%%
  %367 citations counted in INSPIRE as of 20 Jun 2017
	

%\cite{Colangelo-1}
%\cite{Colangelo:2008us}
\bibitem{Colangelo:2008us} 
  P.~Colangelo, F.~De Fazio, F.~Giannuzzi, F.~Jugeau and S.~Nicotri,
  %``Light scalar mesons in the soft-wall model of AdS/QCD,''
  Phys.\ Rev.\ D {\bf 78}, 055009 (2008).
  %doi:10.1103/PhysRevD.78.055009
  %[arXiv:0807.1054 [hep-ph]].
  %%CITATION = doi:10.1103/PhysRevD.78.055009;%%
  %121 citations counted in INSPIRE as of 20 Jun 2017
	
	
%\cite{Vega:2010ne}
\bibitem{Vega:2010ne} 
  A.~Vega and I.~Schmidt,
  %``Modes with variable mass as an alternative in AdS / QCD models with chiral symmetry breaking,''
  Phys.\ Rev.\ D {\bf 82}, 115023 (2010).
  %doi:10.1103/PhysRevD.82.115023
  %[arXiv:1005.3000 [hep-ph]].
  %%CITATION = doi:10.1103/PhysRevD.82.115023;%%
  %34 citations counted in INSPIRE as of 01 Jun 2017
	
	
%\cite{Vega:2011tg}
\bibitem{Vega:2011tg} 
  A.~Vega and I.~Schmidt,
  %``A Chiral symmetry breaking AdS / QCD model with scalar interactions,''
  Phys.\ Rev.\ D {\bf 84}, 017701 (2011).
  %doi:10.1103/PhysRevD.84.017701
  %[arXiv:1104.4365 [hep-ph]].
  %%CITATION = doi:10.1103/PhysRevD.84.017701;%%
  %8 citations counted in INSPIRE as of 01 Jun 2017
	
	
%\cite{Braga:2015jca}
\bibitem{Braga:2015jca} 
  N.~R.~F.~Braga, M.~A.~Martin Contreras and S.~Diles,
  %``Decay constants in soft wall AdS/QCD revisited,''
  Phys.\ Lett.\ B {\bf 763}, 203 (2016).
  %doi:10.1016/j.physletb.2016.10.046
  %[arXiv:1507.04708 [hep-th]].
  %%CITATION = doi:10.1016/j.physletb.2016.10.046;%%
  %3 citations counted in INSPIRE as of 20 Jun 2017
	
%\cite{Braga:2015jcb}
\bibitem{Braga:2015jcb}
    N.~R.~F.~Braga, M.~A.~Martin Contreras and S.~Diles, Eur.\ Phys.\ J. \ C {\bf 76}, 598 (2016).
	
	
%\cite{Vega:2008te}
\bibitem{Vega:2008te} 
  A.~Vega and I.~Schmidt,
  %``Hadrons in AdS/QCD correspondence,''
  Phys.\ Rev.\ D {\bf 79}, 055003 (2009).
  %doi:10.1103/PhysRevD.79.055003
  %[arXiv:0811.4638 [hep-ph]].
  %%CITATION = doi:10.1103/PhysRevD.79.055003;%%
  %45 citations counted in INSPIRE as of 31 May 2017
	
	 
	
%\cite{NIST:DLMF}
\bibitem{NIST:DLMF}
F.~W.~J. Olver, A.~B. {Olde Daalhuis}, D.~W. Lozier, B.~I. Schneider, R.~F. Boisvert, C.~W. Clark, B.~R. Miller and B.~V. Saunders, eds. {\it NIST Digital Library of Mathematical Functions}. \url{http://dlmf.nist.gov/}.


%\cite{Shuryak:1988ck}
\bibitem{Shuryak:1988ck} 
  E.~V.~Shuryak,
  \emph{The QCD vacuum, hadrons and the superdense matter}
  (World Scientific Lecture Notes in Physics, Vol.  {\bf 71}, 1 , Singapore, 2004).
  %[World Sci.\ Lect.\ Notes Phys.\  {\bf 8}, 1 (1988)].
  %%CITATION = 00327,71,1;%%
  %27 citations counted in INSPIRE as of 17 Jan 2017
	
	
%\cite{Cortes:2015emo}
\bibitem{Cortes:2015emo} 
  S.~Cortes, A.~Gomez Nicola and J.~Morales,
  %``Large-$N$ pion scattering at finite temperature: the $f_0(500)$ and chiral restoration,''
  Phys.\ Rev.\ D {\bf 93}, no. 3, 036001 (2016).
  %doi:10.1103/PhysRevD.93.036001
  %[arXiv:1511.00031 [hep-ph]].
  %%CITATION = doi:10.1103/PhysRevD.93.036001;%%
  %3 citations counted in INSPIRE as of 27 Apr 2017
		
	
%\cite{Cortes:2016ecy}
\bibitem{Cortes:2016ecy} 
  S.~Cortes, A.~Gomez Nicola and J.~Morales,
  %``Chiral Symmetry Restoration for the large-$N$ pion gas,''
  Phys.\ Rev.\ D {\bf 94}, no. 11, 116008 (2016).
  %doi:10.1103/PhysRevD.94.116008
  %[arXiv:1609.07751 [hep-ph]].
  %%CITATION = doi:10.1103/PhysRevD.94.116008;%%
  %1 citations counted in INSPIRE as of 27 Apr 2017



\end{thebibliography}
\end{document}